\documentclass[usenatbib,a4]{mn2e}
\usepackage{times}
\usepackage{graphicx}

\pubyear{2008}

\author[Bagla, Prasad and Khandai]
{J. S. Bagla$^1$, Jayanti Prasad$^2$ and Nishikanta Khandai$^3$ \\
  $^1$, $^3$ Harish-Chandra Research Institute,  Chhatnag Road, Jhusi,
  Allahabad 211019, India. \\
  $^2$ National Centre for Radio Astrophysics, Tata Institute of Fundamental
  Research, Post Bag 3, Ganeshkhind, Pune 411007, INDIA \\
  E-mail: $^1$ jasjeet@hri.res.in, $^2$ jayanti@nra.tifr.res.in, $^3$
  nishi@hri.res.in} 
\title[Effects of the size of cosmological N-Body simulations
  on physical quantities]
{Effects of the size of cosmological N-Body simulations
  on physical quantities --- III: Skewness}

\label{firstpage}

\def\LaTeX{L\kern-.36em\raise.3ex\hbox{a}\kern-.15em
    T\kern-.1667em\lower.7ex\hbox{E}\kern-.125emX}

\def\lbx{L_{\rm box}}

\pagerange{\pageref{firstpage}--\pageref{lastpage}}

\begin{document}

\maketitle


\begin{abstract}
N-Body simulations are an important tool in the study of formation of
large scale structures. 
Much of the progress in understanding the physics of galaxy clustering 
and comparison with observations would not have been possible without N-Body
simulations.
Given the importance of this tool, it is essential to understand its
limitations as ignoring these can easily lead to interesting but
unreliable results.
In this paper we study the limitations due to the finite size
of the simulation volume. 
In an earlier work we proposed a formalism for estimating the effects of a
finite box-size on physical quantities and applied it to estimate the effect
on the amplitude of clustering, mass function.  
Here, we extend the same analysis and estimate the effect on skewness and
kurtosis in the perturbative regime. 
We also test the analytical predictions from the earlier work as well as
those presented in this paper.  
We find good agreement between the analytical models and simulations for the
two point correlation function and skewness. 
We also discuss the effect of a finite box size on relative velocity
statistics and find the effects for these quantities scale in a manner that
retains the dependence on the averaged correlation function $\bar\xi$.
\end{abstract}


\begin{keywords}
methods: N-Body simulations, numerical -- gravitation -- cosmology : theory,
dark matter, large scale structure of the universe
\end{keywords}


\section{Introduction}

Large scale structures like galaxies and clusters of galaxies are
believed to have formed by gravitational amplification of small
perturbations.
For an overview and original references, see, e.g.,
\citet{1980lssu.book.....P,1999coph.book.....P,2002tagc.book.....P,2002PhR...367....1B}.  
Density perturbations are present at all scales that have been observed
\citep{2007ApJ...657..645P,2008arXiv0803.0547K}. 
Understanding the evolution of density perturbations for systems that have
fluctuations at all scales is essential for the study of galaxy formation
and large scale structures.  
The equations that describe the evolution of density perturbations in an
expanding universe have been known for a long time \citep{1974A&A....32..391P}
and these are easy to solve when the amplitude of perturbations is
much smaller than unity.   
These equations describe the evolution of density contrast defined as
$\delta(\mathbf r, t) = (\rho(\mathbf r, t) - \bar\rho(t))/\bar\rho(t)$. 
Here $\rho(\mathbf r, t)$ is the density at $\mathbf r$ at time $t$, and
$\bar\rho$ is the average density in the universe at that time.  
These are densities of non-relativistic matter, the component that clusters
at all scales and is believed to drive the formation of large scale structures
in the universe. 
Once the density contrast at relevant scales becomes large, i.e., $|\delta|
\geq 1$, the perturbation becomes non-linear and coupling with perturbations at
other scales cannot be ignored. 
The equations that describe the evolution of density perturbations cannot be
solved for generic perturbations in this regime.
N-Body simulations
\citep{1998ARA&A..36..599B,1997Prama..49..161B,2005CSci...88.1088B,2008SSRv..tmp...26D}
are often used to study the evolution in this regime. 
Alternative approaches can be used if one requires only a limited
amount of information and in such a case either quasi-linear approximation
schemes 
\citep{1970A&A.....5...84Z,1989MNRAS.236..385G,1992MNRAS.259..437M,1993ApJ...418..570B,1994MNRAS.266..227B,1995PhR...262....1S,1996ApJ...471....1H,2002PhR...367....1B}
or scaling relations
\citep{1977ApJS...34..425D,1991ApJ...374L...1H,1995MNRAS.276L..25J,2000ApJ...531...17Ka,1998ApJ...508L...5M,1994MNRAS.271..976N,1996ApJ...466..604P,1994MNRAS.267.1020P,1996MNRAS.278L..29P,1996MNRAS.280L..19P,2003MNRAS.341.1311S}
suffice. 

In cosmological N-Body simulations, we simulate a representative region of the
universe.  
This is a large but finite volume and periodic boundary conditions
are often used.   
Almost always, the simulation volume is taken to be a cube.
Effect of perturbations at scales smaller than the mass resolution of
the simulation, and of perturbations at scales larger than the box is
ignored.
Indeed, even perturbations at scales comparable to the box are under
sampled.

It has been shown that perturbations at small scales do not influence collapse
of perturbations at much larger scales in a significant manner
\citep{1974A&A....32..391P,1985ApJ...297..350P,1991MNRAS.253..295L,1997MNRAS.286.1023B,1998ApJ...497..499C,2008arXiv0802.2796B}. 
This is certainly true if the scales of interest are in the non-linear regime
\citep{1997MNRAS.286.1023B,2008arXiv0802.2796B}. 
Therefore we may assume that ignoring perturbations at scales much smaller
than the scales of interest does not affect results of N-Body simulations. 

Perturbations at scales larger than the simulation volume can affect the
results of N-Body simulations. 
Use of the periodic boundary conditions implies that the average density in
the simulation box is same as the average density in the universe, in other
words we ignore perturbations at the scale of the simulation volume (and at
larger scales).  
Therefore the size of the simulation volume should be chosen so that the
amplitude of fluctuations at the box scale (and at larger scales) is
ignorable.  
If the amplitude of perturbations at larger scales is not ignorable then
clearly the simulation is not a faithful representation of the model
being studied. 
It is not obvious as to when fluctuations at larger scales can be considered
ignorable, indeed the answer to this question depends on the physical quantity
of interest, the model being studied and the specific length/mass scale of
interest as well.  

The effect of a finite box size has been studied using N-Body simulations and
the conclusions in this regard may be summarised as follows.  
\begin{itemize}
\item
If the amplitude of density perturbations around the box scale is small
($\delta < 1$) but not much smaller than unity, simulations underestimate
the correlation function though the number density of small mass haloes does
not change by much \citep{1994ApJ...436..467G,1994ApJ...436..491G}. 
In other words, the formation of small haloes is not disturbed but their
distribution is affected by non-inclusion of long wave modes.
\item
In the same situation, the number density of the most massive haloes drops
significantly
\citep{1994ApJ...436..467G,1994ApJ...436..491G,2005MNRAS.358.1076B}. 
\item
Effects of a finite box size modify values of physical quantities like the
correlation function even at scales much smaller than the simulation volume
\citep{2005MNRAS.358.1076B}.
\item
The void spectrum is also affected by finite size of the simulation volume if
perturbations at large scales are not ignorable \citep{1992ApJ...393..415K}.
\item
It has been shown that properties of a given halo can change significantly as
the contribution of perturbations at large scales is removed to the initial
conditions but the distribution of most internal properties remain unchanged
\citep{2006MNRAS.370..691P}.  
\item
We presented a formalism for estimating the effects of a finite box size in
\citet{2006MNRAS.370..993B}. 
We used the formalism to estimate the effects on the rms amplitude of
fluctuations in density, as well as the two point correlation function.  
We used these to further estimate the effects on the mass function and the
multiplicity function.
\item
The formalism mentioned above was used to estimate changes in the formation
and destruction rates of haloes \citep{2007astro.ph..2557P}.
\item
It was pointed out that the second order perturbation theory and corrections
arising due this can be used to estimate the effects due to a finite box
size \citep{2008MNRAS.389.1675T}.  
This study focused specifically on the effects on baryon acoustic
oscillations. 
\item
If the objects of interest are collapsed haloes that correspond to rare peaks,
as in the study of the early phase of reionisation, we require a fairly large
simulation volume to construct a representative sample of the universe
\citep{2004ApJ...609..474B,2008arXiv0804.0004R}. 
\end{itemize}

In some cases, one may be able to devise a method to ``correct'' for the
effects of a finite box-size \citep{1994A&A...281..301C}, but such methods
cannot be generalised to all statistical measures or physical quantities. 

Effects of a finite box size modify values of physical quantities even at
scales much smaller than the simulation volume
\citep{2005MNRAS.358.1076B,2006MNRAS.370..993B}.  
A workaround for this problem was suggested in the form of an ensemble of
simulations to take the effect of convergence due to long wave modes into
account \citep{2005ApJ...634..728S}, the effects of shear due to long wave
modes are ignored here. 
However it is not clear whether the approach where an ensemble of simulations
is used has significant advantages over using a sufficiently large simulation
volume. 

We review the basic formalism we proposed in \citep{2006MNRAS.370..993B} in
\S{2}. 
We then extend the original formalism to the cases of non-linear amplitude of
clustering and also for estimating changes in skewness and other reduced
moments of counts in cells.
This is done in \S{3}.
In \S{4} we confront our analytical models with N-Body simulations.
We end with a discussion in \S{5}.

\section{The Formalism}

Initial conditions for N-Body simulations are often taken to be a realisation
of a Gaussian random field with a given power spectrum, for details see, e.g.,
\citet{1997Prama..49..161B,1998ARA&A..36..599B,2005CSci...88.1088B,2008SSRv..tmp...26D}.   
The power spectrum is sampled at discrete points in the $\mathbf k$ space
between the scales corresponding to the box size (fundamental mode) and the
grid size (Nyquist frequency/mode).  
Here $\mathbf k$ is the wave vector.  

We illustrate our approach using {\it rms} fluctuations in mass
$\sigma(r)$, but as shown below, the basic approach can be generalised to any
other quantity in a straightforward manner.  
In general, $\sigma(r)$ may be defined as follows:
\begin{equation}
\sigma^2(r) = \int\limits_0^\infty \frac{dk}{k} \frac{k^3 P(k)}{2 \pi^2}
W^2(kr)
\end{equation}
Here $P(k)$ is the power spectrum of density contrast, $r$ is the comoving
length scale at which {\it rms} fluctuations are defined, $k = \sqrt{k_x^2 +
  k_y^2 + k_z^2}$ is the wave number and $W(kr)$ is the Fourier transform of
the window function used for sampling the density field.  
The window function may be a Gaussian or a step function in real or
$k$-space.
We choose to work with a step function in real space where $W(kr) = 9
\left(\sin{kr} - kr \cos{kr}\right)^2/(k^6 r^6)$, see e.g., \S{5.4} of
\citet{1993sfu..book.....P} for further details.    
In an N-Body simulation, the power spectrum is sampled only at specified
points in the $k$-space.  
In this case, we may write $\sigma^2(r)$ as a sum over these points. 
\begin{eqnarray}
\sigma^2(r,\lbx) &=& \frac{9}{V} \sum\limits_{\mathbf k}
P(k) \left[ \frac{\sin{kr} - kr
    \cos{kr}}{k^3 r^3}   \right]^2 \nonumber \\
&\simeq& \int\limits_{2\pi/\lbx}^{2\pi /{L_{\rm grid}}}
\frac{dk}{k} \frac{k^3 P(k)}{2 \pi^2} 9 \left[ \frac{\sin{kr} - kr
    \cos{kr}}{k^3 r^3}   \right]^2 \nonumber \\
&\simeq& \int\limits_{2\pi/\lbx}^\infty
\frac{dk}{k} \frac{k^3 P(k)}{2 \pi^2} 9 \left[ \frac{\sin{kr} - kr
    \cos{kr}}{k^3 r^3}   \right]^2 \nonumber \\
&=& \int\limits_0^\infty
\frac{dk}{k} \frac{k^3 P(k)}{2 \pi^2} 9 \left[ \frac{\sin{kr} - kr
    \cos{kr}}{k^3 r^3}   \right]^2 \nonumber \\
  && ~~~~ - 
\int\limits_0^{2\pi/\lbx}
\frac{dk}{k} \frac{k^3 P(k)}{2 \pi^2} 9 \left[ \frac{\sin{kr} - kr
    \cos{kr}}{k^3 r^3}   \right]^2 \nonumber \\
&=& \sigma_0^2(r) - \sigma_1^2(r,\lbx) 
\end{eqnarray}
Here $\sigma_0^2(r)$ is the
expected level of fluctuations in mass at scale $r$ for the given power
spectrum and $\sigma^2(r,\lbx)$ is what we get in an N-Body simulation at
early times. 
We have assumed that we can approximate the sum over the $k$ modes sampled in
initial conditions by an integral. 
Further, we make use of the fact that small scales do not influence large
scales to ignore the error contributed by the upper limit of the integral. 
This approximation is valid as long as the scales of interest are more than a
few grid lengths. 

In the approach outlined above, the value of $\sigma^2$ at a given scale is
expressed as a combination of the expected value $\sigma_0^2$ and the
correction due to the finite box size $\sigma_1^2$.  
Here $\sigma_0^2$ is independent of the box size and depends only on the power
spectrum and the scale of interest.  
It is clear than $\sigma^2(r,\lbx) \leq \sigma_0^2(r)$, and,
$\sigma_1^2(r,\lbx) \geq 0$.  
It can also be shown that for hierarchical models, $d\sigma_1^2(r,\lbx)/dr
\leq 0$, i.e., $\sigma_1^2(r,\lbx)$ increases or saturates to a constant value
as we approach small $r$. 

At large scales $\sigma_0^2(r)$ and $\sigma_1^2(r,\lbx)$ have a
similar magnitude and the {\it rms} fluctuations in the simulation become
negligible compared to the expected values in the model. 
As we approach small $r$ the correction term $\sigma_1^2(r,\lbx)$ is constant
and for most models it becomes insignificant in comparison with
$\sigma_0^2(r)$.  
In models where $\sigma_0^2(r)$ increases very slowly at small scales or
saturates to a constant value, the correction term $\sigma_1^2$ can be
significant at all scales.  

This formalism can be used to estimate corrections for other estimators of
clustering, for example the two point correlation.   
See \citet{2006MNRAS.370..993B} for details.

The estimation of the {\it rms} amplitude of density perturbations allows us
to use the theory of mass function and estimate a number of quantities of
interest. 
For details, we again refer the reader to \citet{2006MNRAS.370..993B} but we
list important points here.
\begin{itemize}
\item
The fraction of mass in collapsed haloes is under-estimated in N-Body
simulations. 
This under-estimation is most severe near the scale of non-linearity, and
falls off on either side.
If we consider fractional under-estimation in the collapsed fraction then this
increases monotonically from small scales to large scales.
\item
The number density of collapsed haloes is under-estimated at scales larger
than the scale of non-linearity.
The maximum in collapsed fraction near the scale of non-linearity leads to a
change of sign in the effect of a finite box-size for the number density of
haloes at this scale: at smaller scales the number density of haloes is
over-estimated in simulations. 
This can be understood on the basis of a paucity of mergers that otherwise
would have led to formation of high mass haloes.
\item
The above conclusions are generic and do not depend on the specific model for
mass function.  
Indeed, expressions for both the Press-Schechter \citep{1974ApJ...187..425P}
and the Sheth-Tormen \citep{1999MNRAS.308..119S,2001MNRAS.323....1S} mass
functions are given in \citet{2006MNRAS.370..993B}, and we have also checked
the veracity of our claims for the \citet{2001MNRAS.321..372J} mass function.
\end{itemize}

\section{Reduced Moments}

In this section we outline how the formalism and results outlined above may be
used to estimate the effect of a finite box-size on reduced moments.
Reduced moments like the skewness and kurtosis can be computed using
perturbation theory in the weakly non-linear regime
\citep{1994ApJ...433....1B}.  
The expected values of the reduced moments are related primarily to the slope
of the initial or linearly extrapolated $\sigma^2(r)$, as all
non-Gaussianities are generated through evolution of the Gaussian initial
conditions and the initial $\sigma^2(r)$ characterises this completely. 
We can use the expression for $\sigma^2(r)$ as it is realised in simulations
with a finite box size to compute the expected values of reduced moments in
N-Body simulations in the weakly non-linear regime.
\begin{eqnarray}
S_3 &=& \frac{34}{7} + \frac{\partial\ln\sigma^2}{\partial\ln r} \nonumber \\
&=& \frac{34}{7} +
\frac{\partial\ln\left(\sigma_0^2-\sigma_1^2\right)}{\partial\ln r} \nonumber
\\
&=& \frac{34}{7} + \frac{\partial\ln\sigma_0^2}{\partial\ln r} +
\frac{\partial\ln\left(1-\sigma_1^2/\sigma_0^2\right)}{\partial\ln r}
\nonumber \\ 
&=& S_{3_0}-S_{3_1}
\end{eqnarray}
$S_{3_0}$ is the expected value of $S_3$ for the given mode, i.e., when there
are no box corrections and $S_{3_1}$ is the correction term in $S_3$ due to a
finite box size.
Box size effects lead to a change in slope of $\sigma^2$, and hence the
effective value of $n$ changes.
The last term is the offset in skewness in N-Body simulations as
compared with the expected values in the model being simulated. 
We would like to emphasise that this expression is valid only in the weakly
non-linear regime.

In general we expect $\sigma_1^2/\sigma_0^2$ to increase as we go to larger
scales. 
Thus the skewness is under estimated in N-Body simulations and the level of
under estimation depends on the slope of $\sigma_1^2/\sigma_0^2$ as compared
to the slope of $\sigma_0^2$. 
In the limit of small scales where $\sigma_1^2$ is almost independent of
scale, we find that the correction is:
\begin{eqnarray}
S_3 &=&\frac{34}{7} + \frac{\partial\ln\sigma_0^2}{\partial\ln r} +
\frac{\partial\ln\left(1-\sigma_1^2/\sigma_0^2\right)}{\partial\ln r}
\nonumber\\
&\simeq& \frac{34}{7} - \left(n+3\right) -
\frac{\partial\left({\sigma_1^2}/{\sigma_0^2}\right)}{\partial\ln{r}} +
\mathcal{O}\left(\frac{\partial}{\partial\ln{r}}\left(
    \frac{\sigma_1^2}{\sigma_0^2} \right)^2 \right) 
\nonumber \\
&\simeq& \frac{34}{7} - \left(n+3\right) \left[ 1 +
  \frac{\sigma_1^2}{\sigma_0^2} \right] +
\mathcal{O}\left(
  \frac{\partial}{\partial\ln{r}}\left(\frac{\sigma_1^2}{\sigma_0^2} 
  \right)^2 \right) \nonumber \\  
&& +
\mathcal{O}\left(\frac{1}{\sigma_0^2}
  \frac{\partial\sigma_1^2}{\partial\ln{r}} \right)    
\end{eqnarray}
Here $n$ is the index of the initial spectrum we are simulating. 
For non-power law models this will also be a function of scale.
The correction becomes more significant at larger scales and the net effect,
as noted above, is to under estimate $S_3$.

Similar expressions can be written down for kurtosis and other reduced
moments using the approach outlined above. 
We give the expression for kurtosis below, but do not compute further
moments as the same general principle can be used to compute these as well. 
\begin{eqnarray}
S_4 &=& \frac{6071}{1323} + \frac{62}{3}
\frac{\partial\ln\sigma^2}{\partial\ln r} \nonumber \\
&& + \frac{7}{3} \left[ \frac{\partial\ln\sigma^2}{\partial\ln r} \right]^2
+ 2 \frac{\partial^2\ln\sigma^2}{\partial\ln^2 r} \nonumber \\
&\simeq& \frac{6071}{1323} - \frac{62}{3}(n+3) \left[1 +
  \frac{\sigma_1^2}{\sigma_0 ^2} \right]  + \frac{7}{3} (n+3)^2 \left[1 -
  \frac{8}{7} \frac{\sigma_1^2}{\sigma_0^2} \right]     \nonumber \\
&&   +
\mathcal{O}\left(
  \frac{\partial}{\partial\ln{r}}\left(\frac{\sigma_1^2}{\sigma_0^2} 
  \right)^2 \right) + 
\mathcal{O}\left(\frac{1}{\sigma_0^2}
  \frac{\partial\sigma_1^2}{\partial\ln{r}} \right)  
\end{eqnarray}

\section{N-Body Simulations}

In this section we compare the analytical estimates for finite box size
effects for various quantities with N-Body simulations. 
Such a comparison is relevant in order to test the effectiveness of
approximations made in computing the effects of a finite box size. 
We have made the following approximations:
\begin{itemize}
\item
Effects of mode coupling between the scales that are taken into account in a
simulation and the modes corresponding to scales larger than the simulation
box are ignored.
We believe that this should not be important unless the initial power spectrum
has a sharp feature at scales comparable with the simulation size\footnote{For
  example, simulations of baryon acoustic oscillations imprinted in the matter
  power spectrum may be affected by mode coupling even though the amplitude of
  fluctuations at the relevant scales is very small
  \citep{1974A&A....32..391P,1997MNRAS.286.1023B,2008MNRAS.389.1675T}.}.  
\item
Sampling of modes comparable to the box size is sparse, and the approximation
of the sum over wave modes as an integral can be poor if the relative
contribution of these scales to $\sigma_1$ is significant.  
\end{itemize}

Table~1 gives details of the N-Body simulations used in this paper.  
In order to simulate the effects of a finite box size, we used the method
employed by \citet{2005MNRAS.358.1076B} where initial perturbations are set to
zero for all modes with wave number smaller than a given cutoff $k_c$. 
The initial conditions are exactly the same as the reference simulation in
each series in all other respects. 
For a finite simulation box, there is a natural cutoff at the fundamental wave
number $k_f=2\pi/\lbx$ and simulations A1, B1 and C1 impose no other cutoff. 
These are the reference simulations for the two series of simulations. 
Simulations A2, B2 and C2 sample perturbations at wave numbers larger than
$2k_f$ whereas simulations A3, B3 and C3 are more restrictive with non-zero
perturbations above $4k_f$.  
The cutoff of $2k_f$ and $4k_f$ corresponds to scales of $128$ and $64$ grid
lengths, respectively. 
For the C series of simulations, the cutoff of $2k_f$ and $4k_f$ corresponds
to scales of $80$~h$^{-1}$Mpc and $40$~h$^{-1}$Mpc, respectively.

The background cosmology was taken to be Einstein-deSitter for the A and B
series simulations.
The best fit $\Lambda$CDM model from WMAP-5 \citep{2008arXiv0803.0586D} was
used for the C series of simulations. 

In order to ensure that the initial conditions do not get a rare contribution
from a large scale mode, we forced $|\delta_k|^2 = P(k)$ while keeping the
phases random for modes $k \geq 6  k_f$.  

We have chosen to work with models where box size effects are likely to be
significant, particularly with the larger cutoff in wave number.
This has been done to test our analytical model in a severe situation, and
also to further illustrate the difficulties in simulating models with large
negative indices.

We present results from N-Body simulations in the following section.

\begin{table}
\begin{center}
\begin{tabular}{||c|c|c||} \hline \hline
Model    & Description  & Cut Off  ($k_c$)    \\ \hline \hline
A1       & Power Law, $n-2.0$ & $k_f$            \\ \hline
A2       & Power Law, $n-2.0$ & $2k_f$       \\ \hline
A3       & Power Law, $n-2.0$ & $4k_f$    \\ \hline
B1       & Power Law, $n-2.5$ & $k_f$      \\ \hline
B2       & Power Law, $n-2.5$ & $2k_f$     \\ \hline
B3       & Power Law, $n-2.5$ & $4k_f$  \\ \hline 
C1       & $\Lambda$CDM, WMAP-5 BF, $L_{box}=160$~h$^{-1}$Mpc & $k_f$      \\ \hline
C2       & $\Lambda$CDM, WMAP-5 BF, $L_{box}=160$~h$^{-1}$Mpc & $2k_f$     \\ \hline
C3       & $\Lambda$CDM, WMAP-5 BF, $L_{box}=160$~h$^{-1}$Mpc & $4k_f$  \\ \hline \hline
\end{tabular}
\caption{This table lists characteristics of N-Body simulations used in our
  study.  Here the spectral index gives the slope of the initial power
  spectrum and the cutoff refers to the wave number below which all
  perturbations are set to zero: $k_f=2\pi/\lbx$ is the fundamental wave
  mode for the simulation box.  All models were simulated using the TreePM
  code \citep{2002JApA...23..185B,2003NewA....8..665B,2008arXiv0802.3215K}.
  $256^3$ particles were used in each simulation, and the PM
  calculations were done on a $256^3$ grid.  Power spectra for both the A and
  the B series of simulations were normalised to ensure $\sigma=1$ at the 
  scale of $8$ grid lengths at the final epoch if there is no box-size
  cutoff.  A softening length of $0.1$ grid lengths was used as the evolution 
  of small scale features is not of interest in the present study.
  Simulations for both the A and the B series were done with the
  Einstein-deSitter background and the C series used the WMAP-5 best fit (BF)
  model as the cosmological background, as also for the power spectrum and
  transfer function.} 
\end{center}
\end{table}

\begin{figure*}
\begin{center}
\begin{tabular}{cc}
\includegraphics[width=2.8truein]{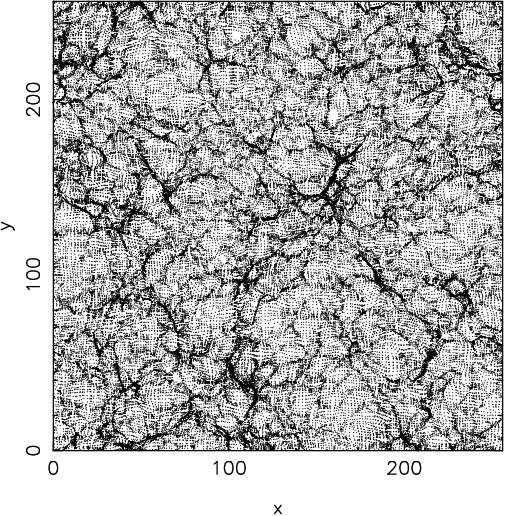} &
\includegraphics[width=2.8truein]{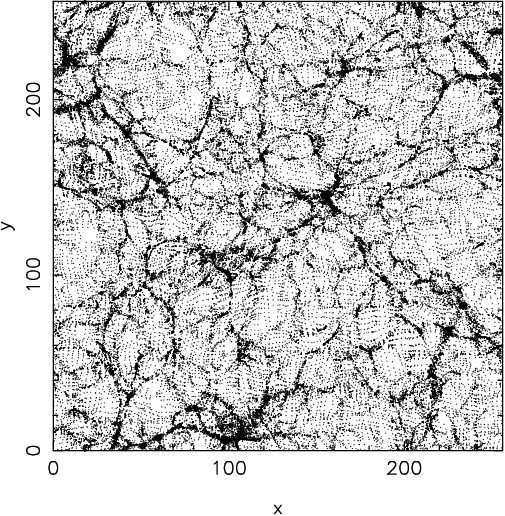} \\
\includegraphics[width=2.8truein]{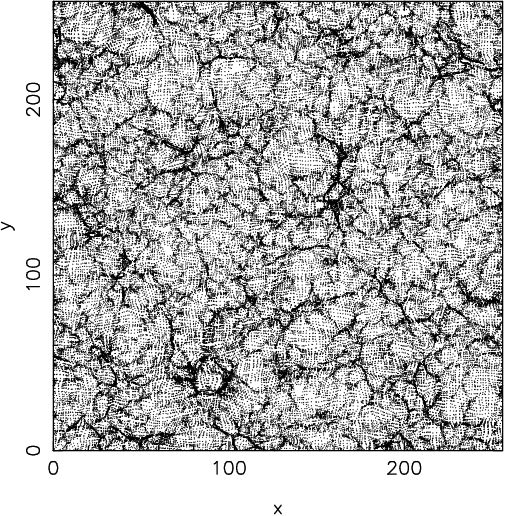} &
\includegraphics[width=2.8truein]{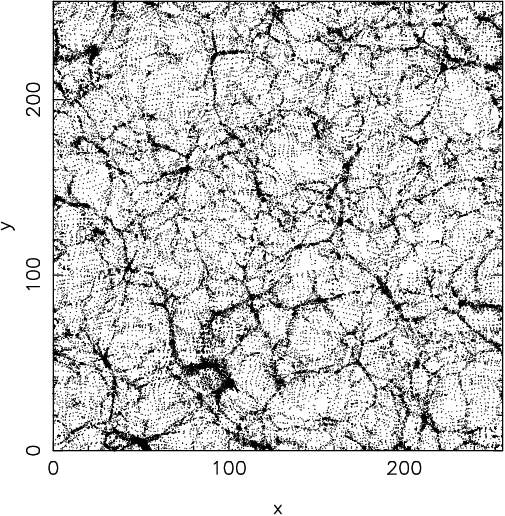} \\
\includegraphics[width=2.8truein]{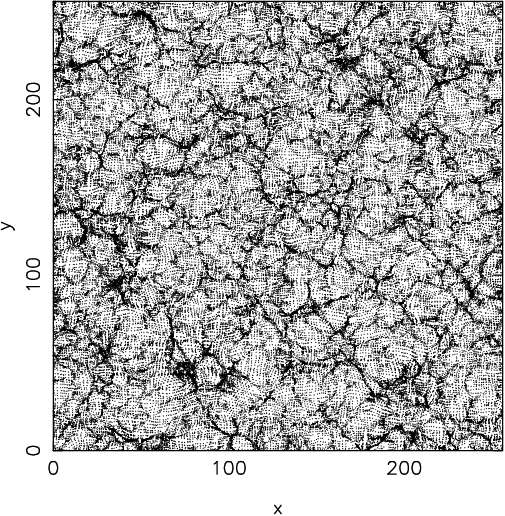} &
\includegraphics[width=2.8truein]{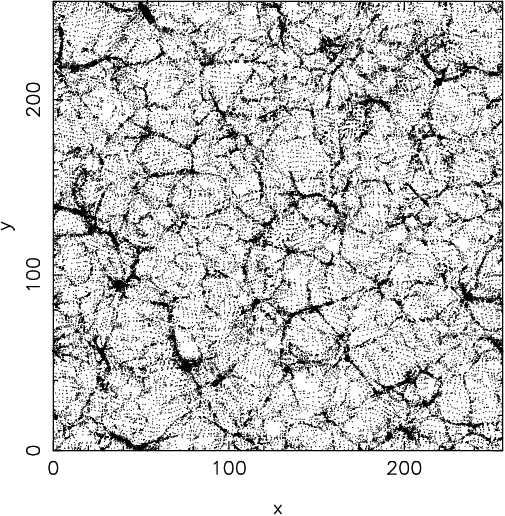}
\end{tabular}
\end{center}
\caption{The first, second and the third row in this figure show the 
slices for  models A1, A2 and A3 (see table for details) respectively 
at an early epoch when the scale of nonlinearity is $2$ grid lengths 
(left column) and a later epoch when the scale of nonlinearity is $8$ 
grid lengths (right column)}. 
\end{figure*}

\begin{figure*}
\begin{center}
\begin{tabular}{cc}
\includegraphics[width=2.8truein]{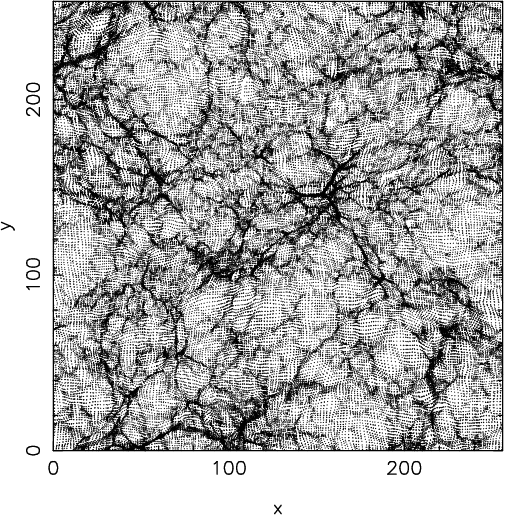} &
\includegraphics[width=2.8truein]{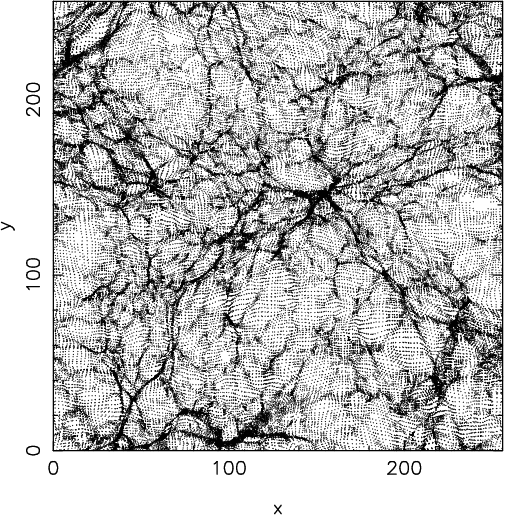} \\
\includegraphics[width=2.8truein]{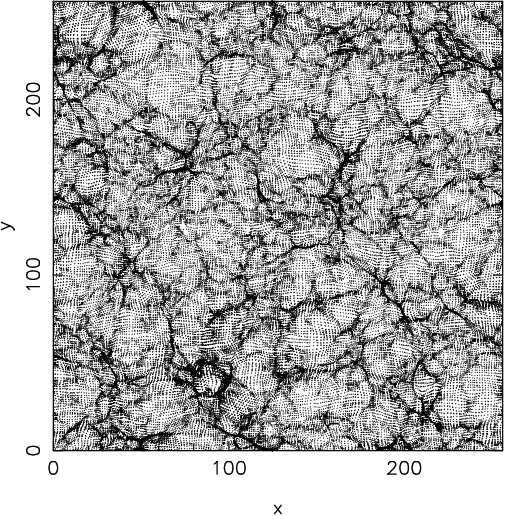} &
\includegraphics[width=2.8truein]{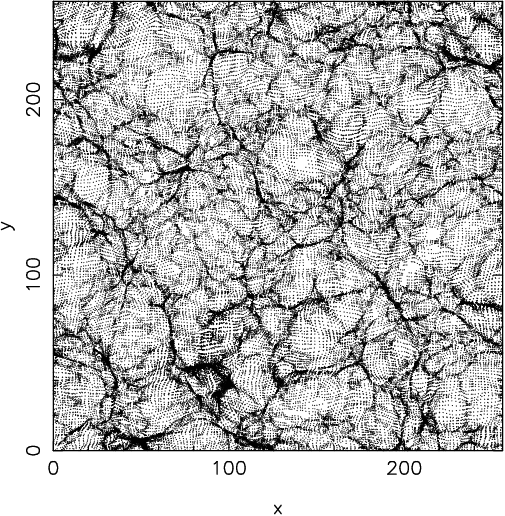} \\
\includegraphics[width=2.8truein]{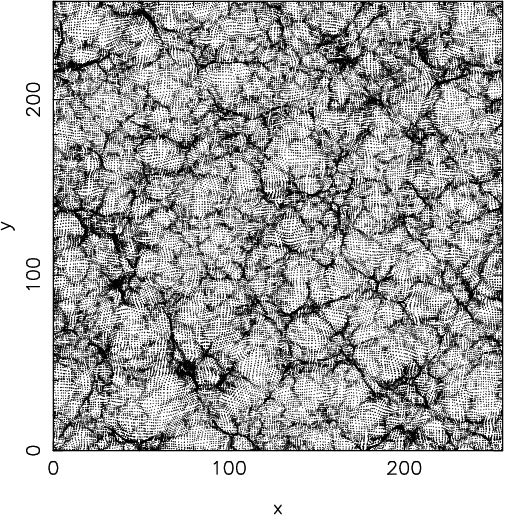} &
\includegraphics[width=2.8truein]{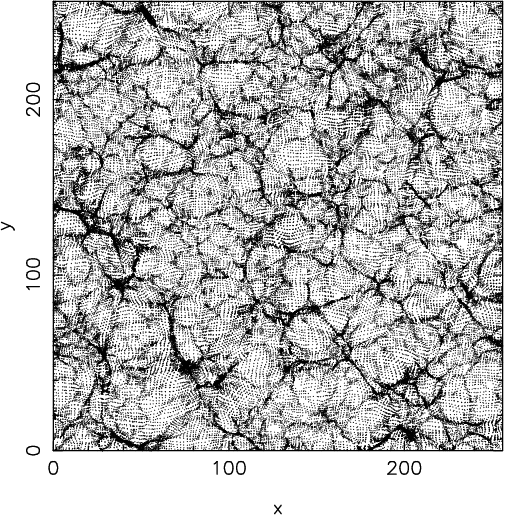}
\end{tabular}
\end{center}
\caption{The first, second and the third row in this figure show the
slices for the models B1, B2 and B3 respectively at an early epoch
when the scale of nonlinearity is $2$ grid lengths (left column) and a 
later epoch (right column) when the scale of nonlinearity is $8$ grid 
lengths (right column)}.
\end{figure*}

\begin{figure*}
\begin{center}
\begin{tabular}{cc}
\includegraphics[width=2.8truein]{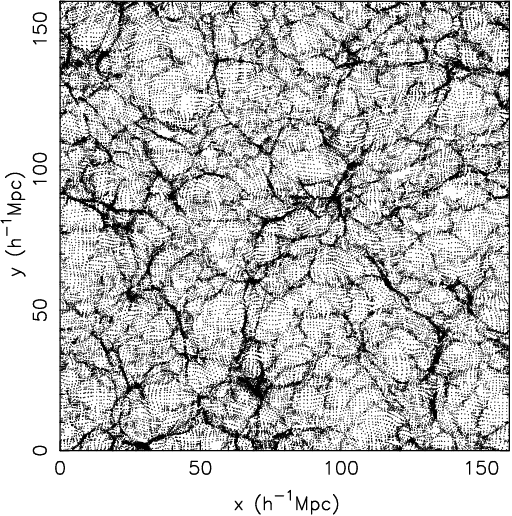} &
\includegraphics[width=2.8truein]{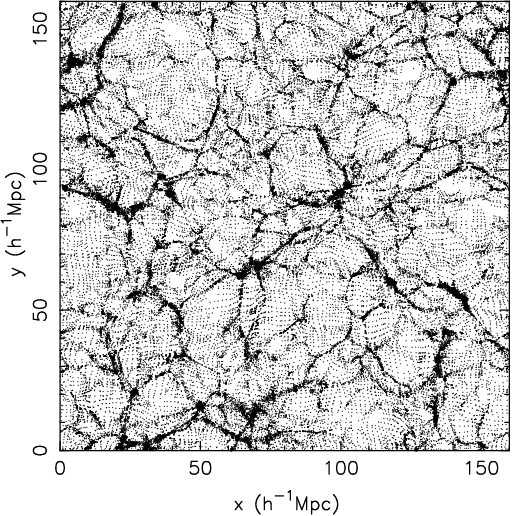} \\
\includegraphics[width=2.8truein]{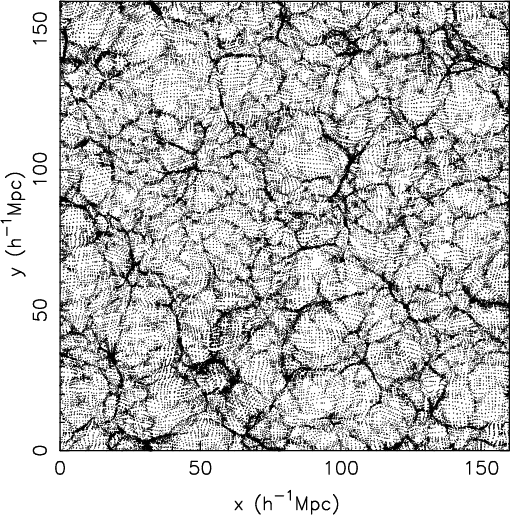} &
\includegraphics[width=2.8truein]{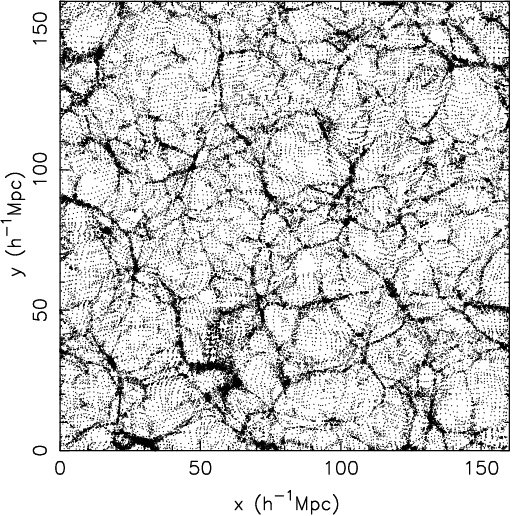} \\
\includegraphics[width=2.8truein]{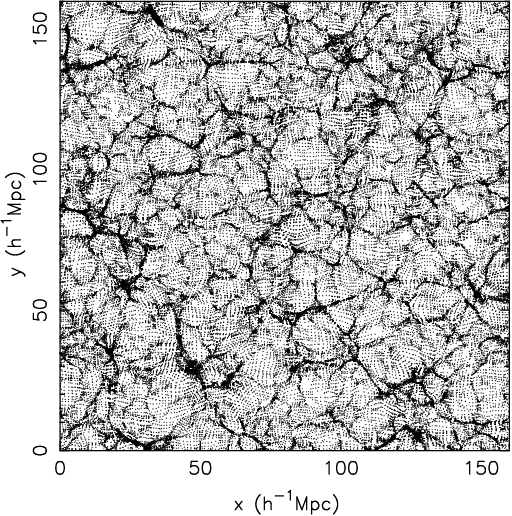} &
\includegraphics[width=2.8truein]{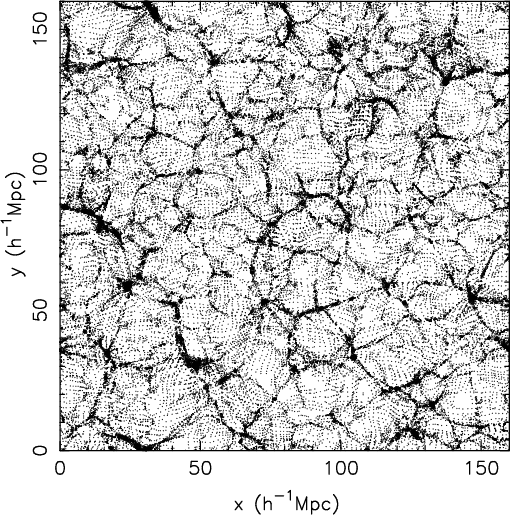}
\end{tabular}
\end{center}
\caption{The first, second and the third row in this figure show the
slices for the $\Lambda$CDM simulations C1, C2 and C3 respectively at an early
epoch ($z=1$) (left column) and the present epoch ($z=0$) (right column).}
\end{figure*}

\begin{figure*}
\begin{center}
\begin{tabular}{ccc}
\includegraphics[height=2.1truein]{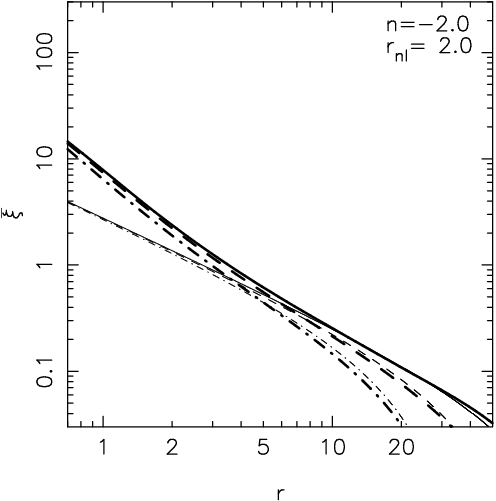} &
\includegraphics[height=2.1truein]{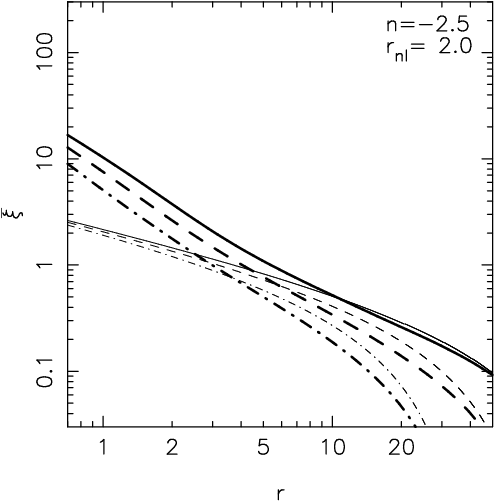} &
\includegraphics[height=2.1truein]{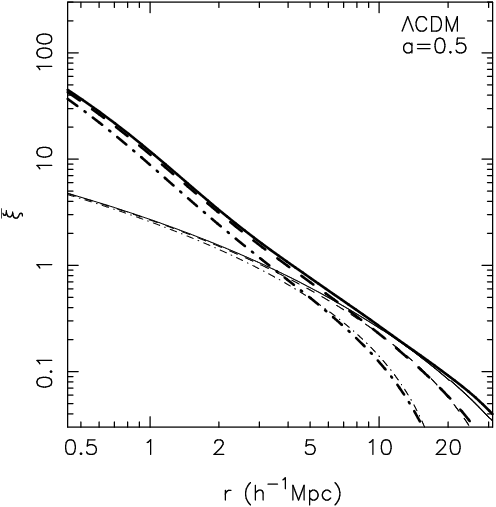} \\
\includegraphics[height=2.1truein]{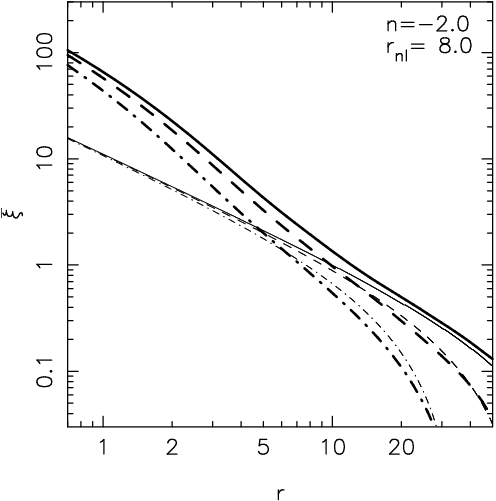} &
\includegraphics[height=2.1truein]{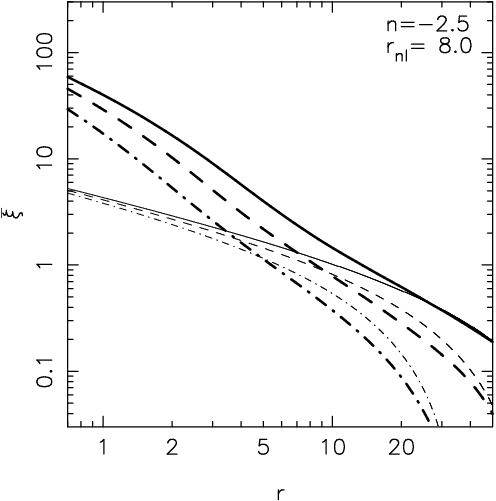} &
\includegraphics[height=2.1truein]{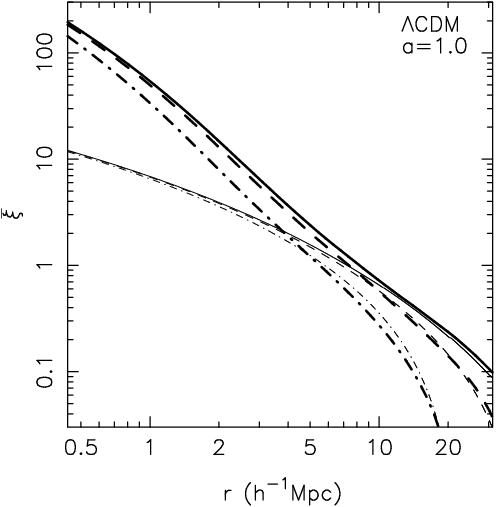} \\
\includegraphics[height=2.1truein]{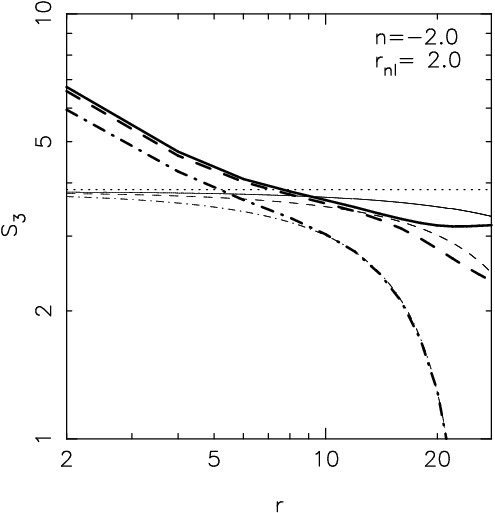} &
\includegraphics[height=2.1truein]{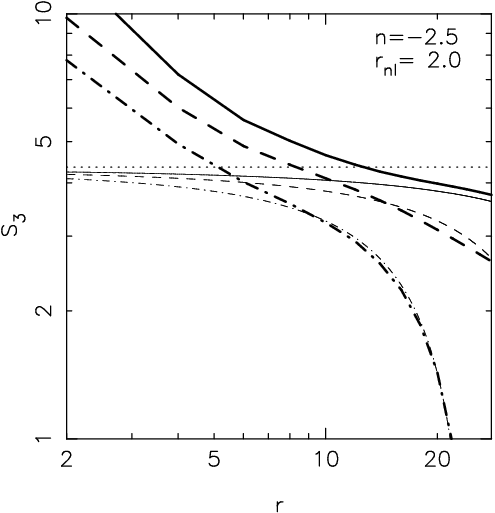} &
\includegraphics[height=2.1truein]{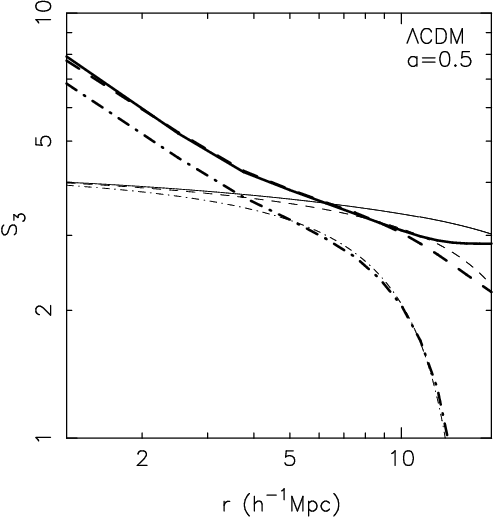} \\
\includegraphics[height=2.1truein]{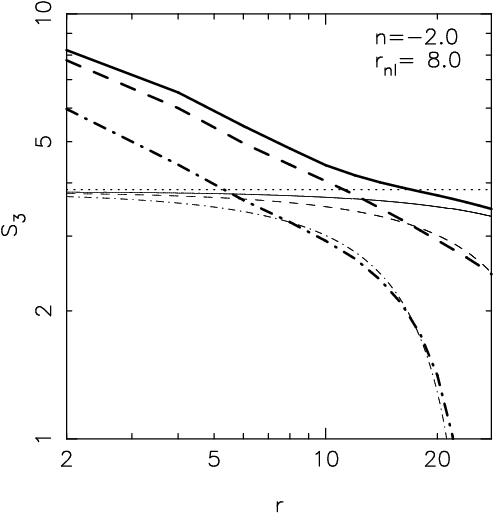} &
\includegraphics[height=2.1truein]{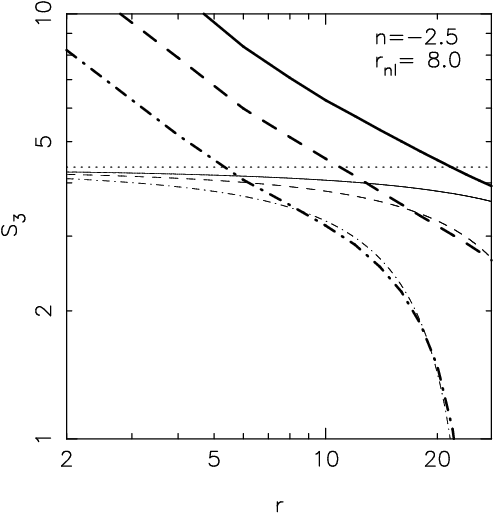} &
\includegraphics[height=2.1truein]{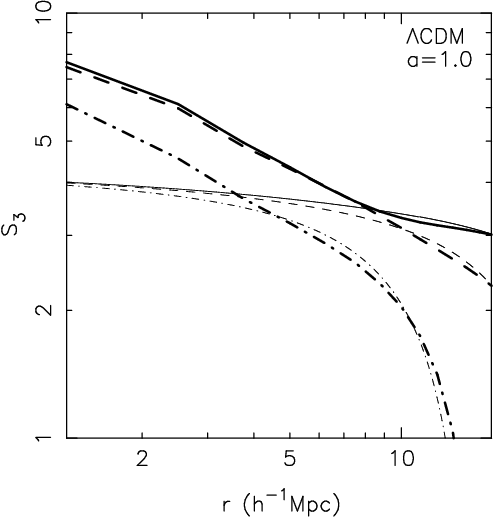} \\
\end{tabular}
\end{center}
\caption{The first two rows in this figure show the
average two point correlation function ${\bar \xi}$, and the next two rows
show skewness $S_3$.
The first and third row represent the early epoch and the second and fourth 
row represent the later epoch respectively.
In all the panels models with $k_c=k_f$, $k_c=2k_f$ and $k_c=4k_f$ are
represented by the solid,  dashed and dot-dashed lines respectively.
In all the  panels,  corresponding to every model in simulation (thick lines)
theoretical estimates (thin lines) are also shown. 
Horizontal dashed lines in the lower rows shows the expected value of $S_3$ in
absence of any box-size corrections for the power law models.
}
\end{figure*}

\begin{figure*}
\begin{center}
\begin{tabular}{ccc}
\includegraphics[height=2.1truein]{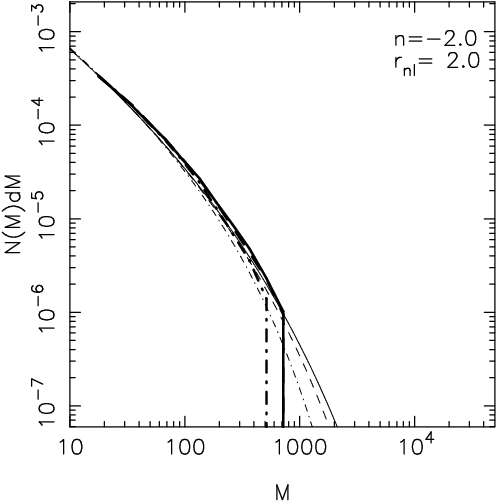} &
\includegraphics[height=2.1truein]{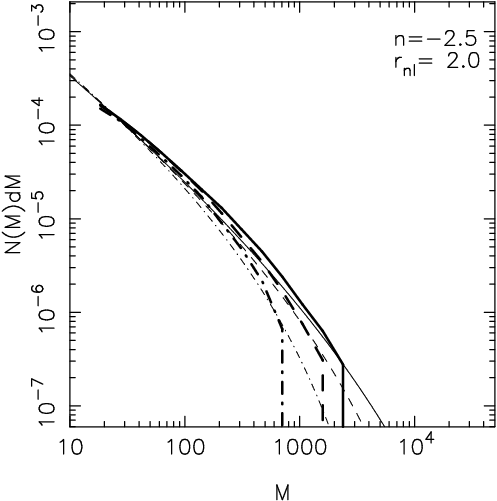} &
\includegraphics[height=2.1truein]{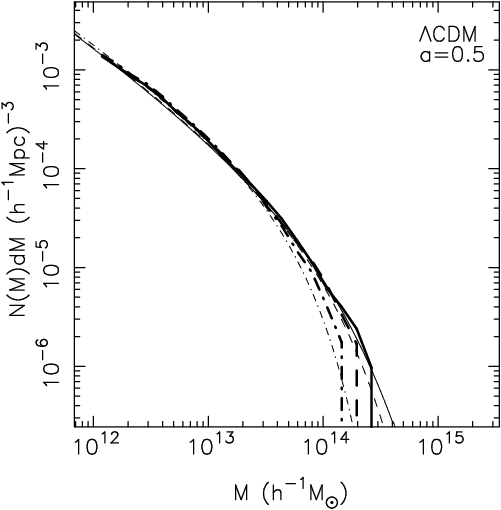} \\
\includegraphics[height=2.1truein]{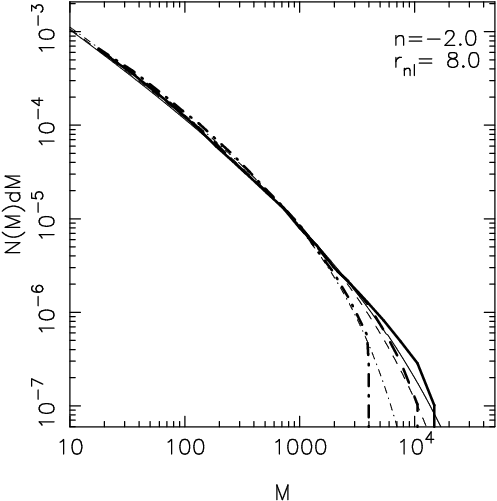} &
\includegraphics[height=2.1truein]{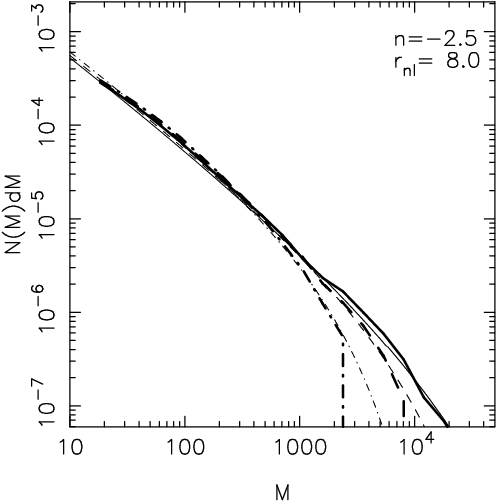} &
\includegraphics[height=2.1truein]{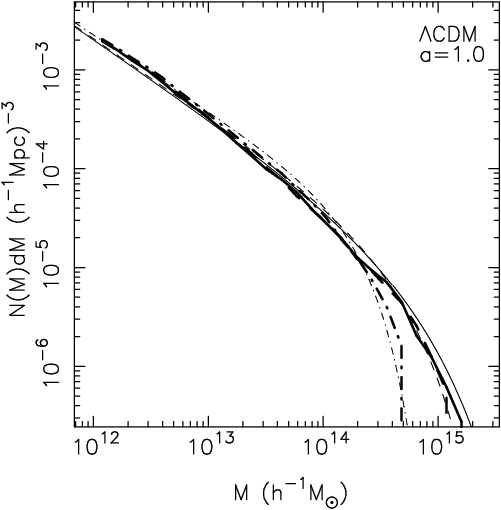} \\
\end{tabular}
\end{center}
\caption{This figure shows the mass function $N(M)dM$ for the three series of
  simulations.  
The top row shows this for the early epoch and the lower row corresponds to
the late epoch.
In all the panels models with $k_c=k_f$, $k_c=2k_f$ and $k_c=4k_f$ are
represented by the solid,  dashed and dot-dashed lines respectively.
In all the  panels,  corresponding to every model in simulation (thick lines)
theoretical estimates (thin lines) are also shown.
}
\end{figure*}

\subsection{Results}

We begin with a visual representation of the simulations.  
Figure~1 shows a slice from simulations A1, A2 and A3 at two different
epochs.  
The left panel is for the early epoch when $r_{nl}=2$~grid lengths in the
model without a cutoff, and the right panel is for $r_{nl}=8$~grid lengths. 
The top row is for the simulation A1, the middle row is for the simulation A2
and the lowest row is for the simulation A3.
The relevance of box size effects is apparent as the large scale structure in
the three simulations is very different even at the early epoch when
$r_{nl}=2$, much smaller than the effective box size for these simulations. 
Disagreement between different simulations becomes even more severe as we go
to the later epoch with $r_{nl}=8$~grid lengths.  

Visual appearance for simulations B1, B2 and B3, shown in Figure~2 follows the
same pattern.  
In this case the spectral index is closer to $-3$ than for simulations of the
A series shown in Figure~1, hence the larger scale modes are more important
for evolution of perturbations even at small scales. 
It is interesting to note that the largest under-dense region in simulation B1
at early times is already comparable to the box size and hence we require
$\lbx/r_{nl} \gg 128$ for the effects of a finite box-size to be small
enough to be ignored for simulations of the power law model with $n=-2.5$.
This constraint is even stronger for models with the slope of the power
spectrum closer to $n=-3$. 

Figure~3 shows the visual appearance for the C series of simulations.  
Once again we find a significant change in appearance even with $k_c=2 k_f$ at
the earlier epoch, $z=1$ in this case. 
This indicates that a box-size of $80$~h$^{-1}$Mpc is insufficient if we wish
to achieve convergence in the large scale distribution of matter in models of
cosmological interest.
This reinforces conclusions of \citet{2005MNRAS.358.1076B} where we found that
a box size of around $150$~h$^{-1}$Mpc is required for convergence in
simulations of $\Lambda$CDM models.

\subsection{Clustering Amplitude}

The left column in Figure~4 shows the volume averaged correlation function
$\bar\xi$ for the simulations being studied here.  
The top-left panel is for simulations A1, A2 \& A3 at an early epoch
($r_{nl}=2$ grid lengths in the model without a cutoff) and the second panel
from top in this column shows $\bar\xi$ for the same simulations at a late
epoch ($r_{nl}=8$ grid lengths.   
The corresponding plots in the second column show the same for the simulations
in the B series, and the third column is for the C series of simulations. 
We have shown $\bar\xi$ as a function of scale in these panels. 
Also shown are the linearly extrapolated values of $\bar\xi$ computed using
our formalism for estimating the effects of a finite box-size. 
Data from N-Body simulations is shown as thick curves whereas the theoretical
estimate is shown as thin curves with the corresponding line style. 
It is clear that the analytical estimate for $\bar\xi$ in a finite box
captures the qualitative nature of the change from the expected values.
The match is better at large scales where $\bar\xi$ is small and this is
expected as the analytical estimate is linearly extrapolated whereas we are
comparing it with results from an N-Body simulation.   
Our analysis works better for the $n=-2$ model used in the A series of
simulations and for the $\Lambda$CDM model in the C series of simulations as
compared to the B series of simulations for the $n=-2.5$ model 
where it systematically under-estimates the suppression of $\bar\xi$. 
It is noteworthy that even in this case the differences between the simulation
and the analytical model at large scales is of order of $20-30\%$ whereas the
box-size effect changes the clustering amplitude by more than an order of
magnitude at some scales. 
Thus we may state that the model captures the essence of the box-size effects
at large scales.

\subsection{Skewness}

The lower two rows in Figure~4 show the corresponding plots for $S_3$, shown
here as a function of scale. 
Apart from the lines that show $S_3$ from simulations (thick lines) and our
analytical estimate for the weakly non-linear regime (thin lines), we also
show the value of $S_3$ expected in the weakly non-linear regime in absence of
any finite box size effects for the three series of simulations. 
The analytical estimate of $S_3$, computed using Eqn.(3) matches well with the
values in N-Body simulation at large scales. 
It is noteworthy that the match between the two is better for a larger cutoff
in wave numbers. 
We believe that this is due to sparse sampling of the initial power spectrum
at scales comparable to the box size and due to this our approximation of the
sum over wave modes by an integral is not very good. 

\begin{figure*}
\begin{center}
\begin{tabular}{ccc}
\includegraphics[height=2.1truein]{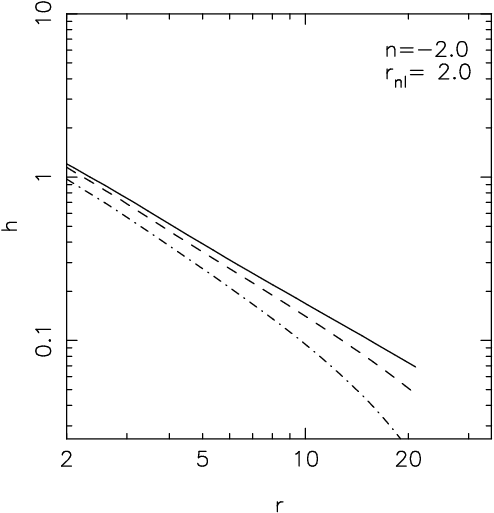} &
\includegraphics[height=2.1truein]{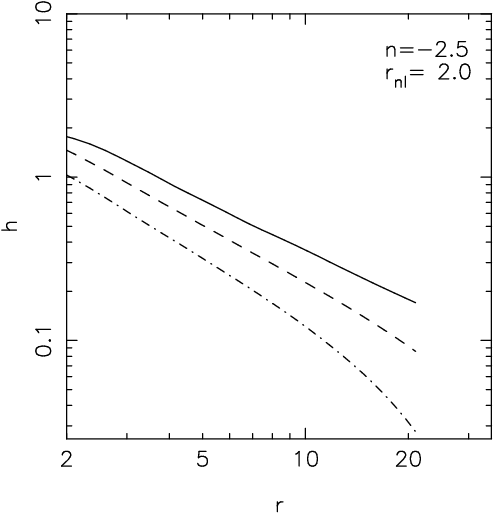} &
\includegraphics[height=2.1truein]{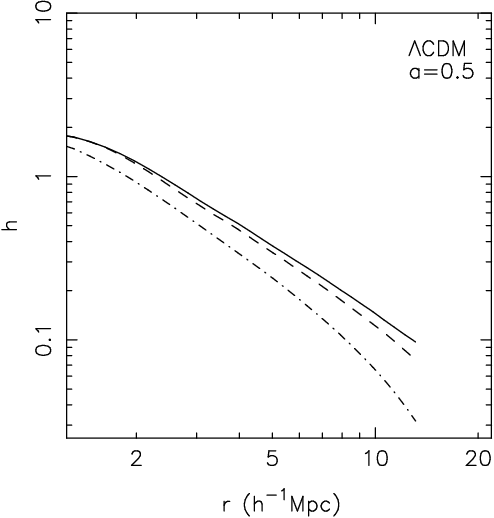} \\
\includegraphics[height=2.1truein]{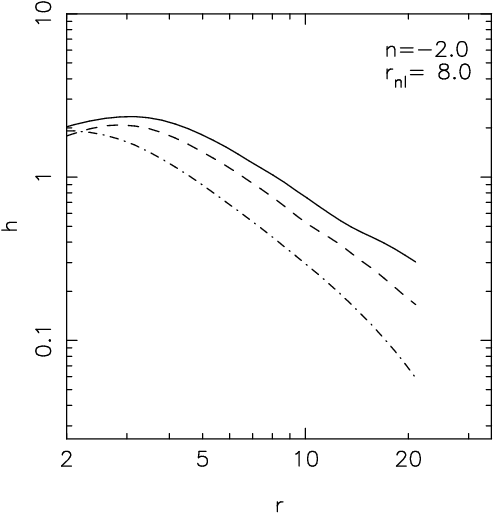} &
\includegraphics[height=2.1truein]{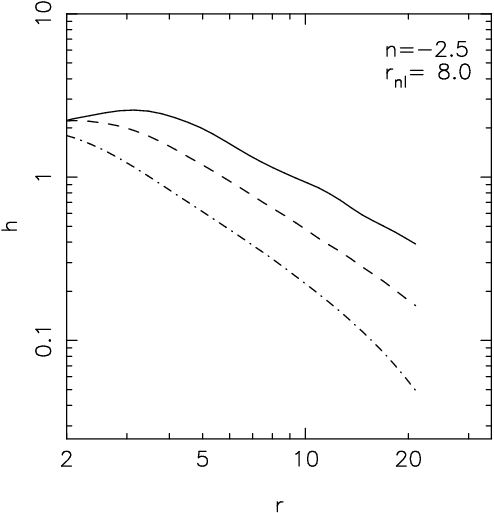} &
\includegraphics[height=2.1truein]{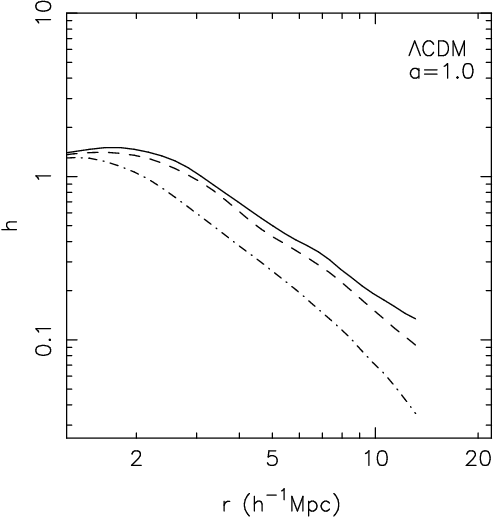} \\
\includegraphics[height=2.1truein]{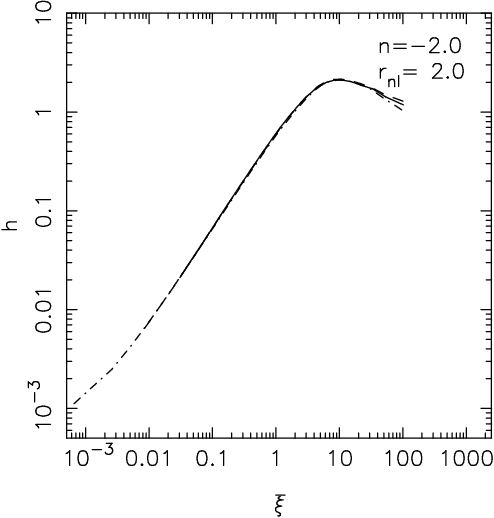} &
\includegraphics[height=2.1truein]{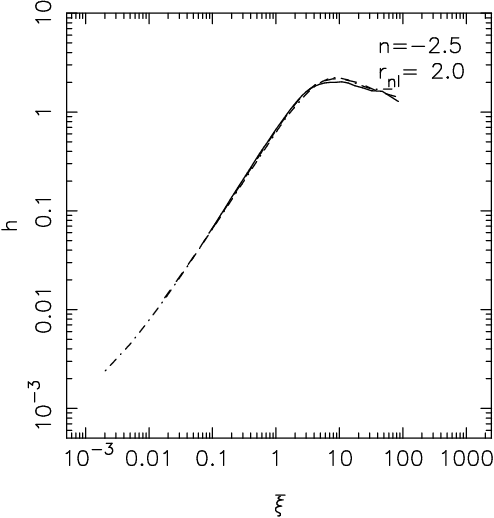} &
\includegraphics[height=2.1truein]{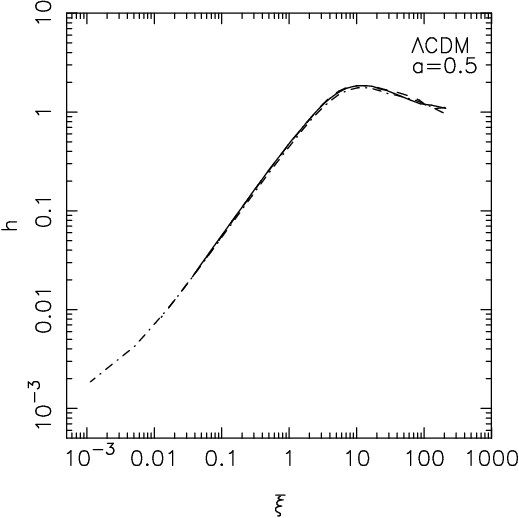} \\
\includegraphics[height=2.1truein]{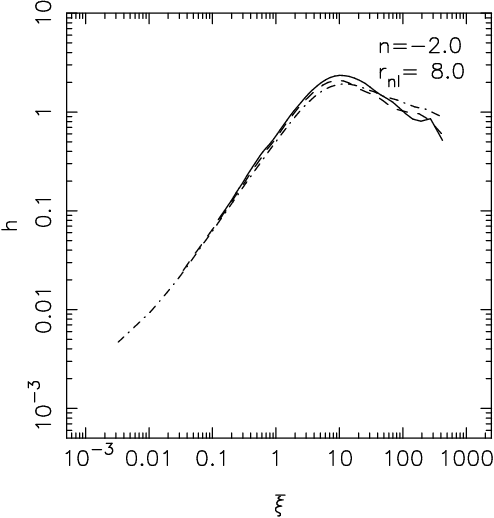} &
\includegraphics[height=2.1truein]{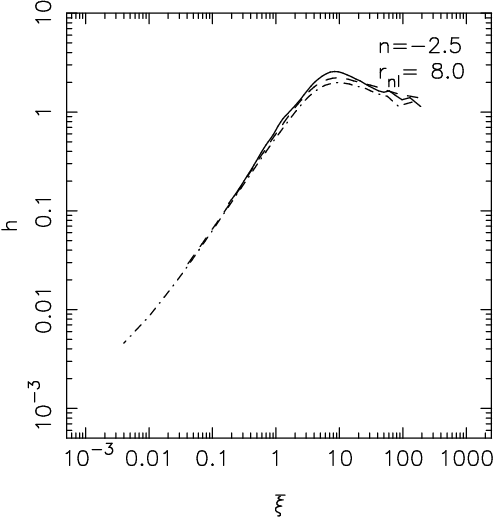} &
\includegraphics[height=2.1truein]{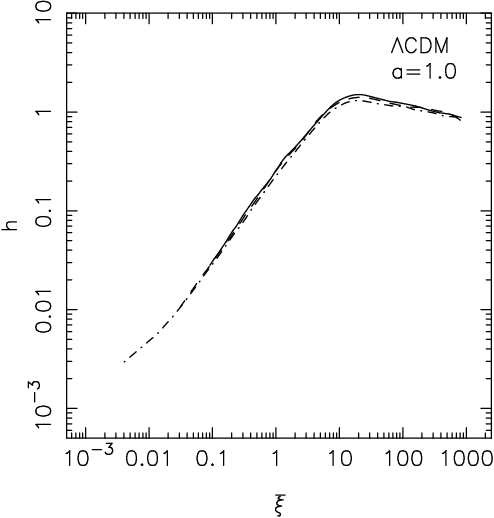} \\
\end{tabular}
\end{center}
\caption{The first two rows show the pair velocity as a function of distance,
  and the lowest two rows show the pair velocity as a function of average two
  point correlation function.
The first and third row represent the early epoch and 
second and fourth row represent the later epoch.
In all panels models with $k_c=k_f$ (A1, B1 and C1), $k_c=2k_f$ (A2, B2 and
  C2) and $k_c=4k_f$ (A3, B3 and C3) are represented by the solid, dashed and 
dot-dashed lines respectively.
}
\end{figure*}
\begin{figure*}
\begin{center}
\begin{tabular}{ccc}
\includegraphics[height=2.1truein]{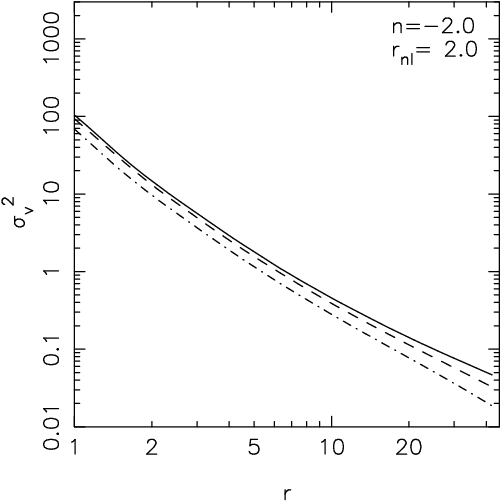} &
\includegraphics[height=2.1truein]{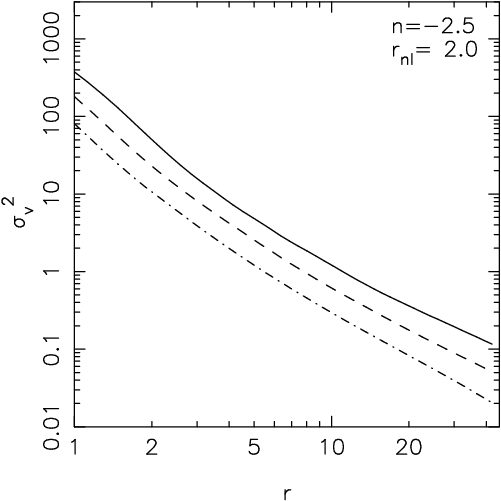} &
\includegraphics[height=2.1truein]{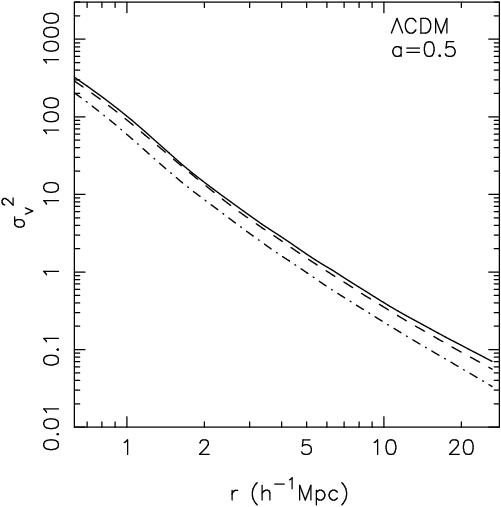} \\
\includegraphics[height=2.1truein]{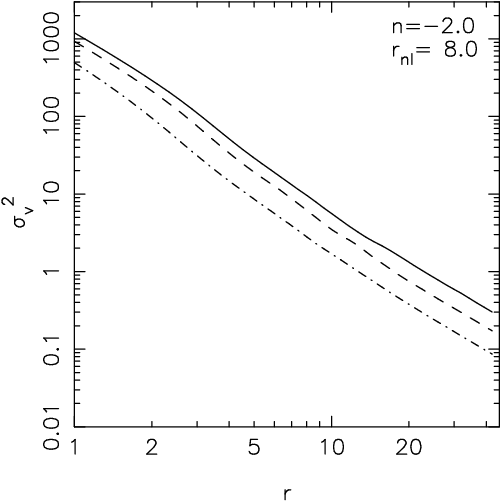} &
\includegraphics[height=2.1truein]{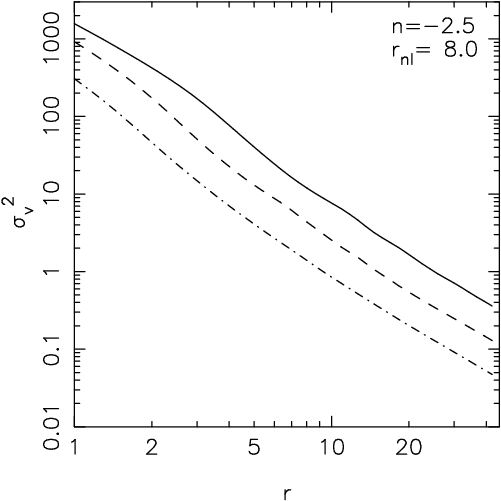} &
\includegraphics[height=2.1truein]{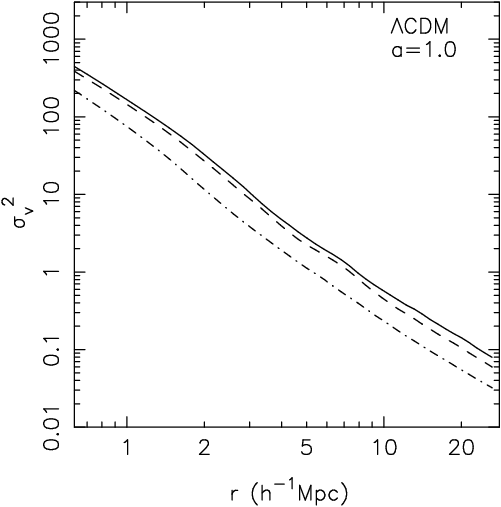} \\
\includegraphics[height=2.1truein]{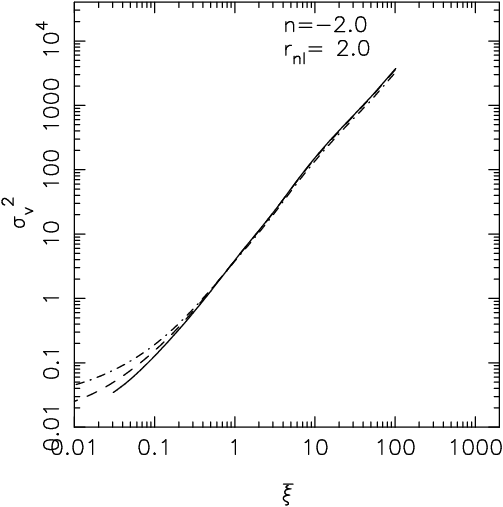} &
\includegraphics[height=2.1truein]{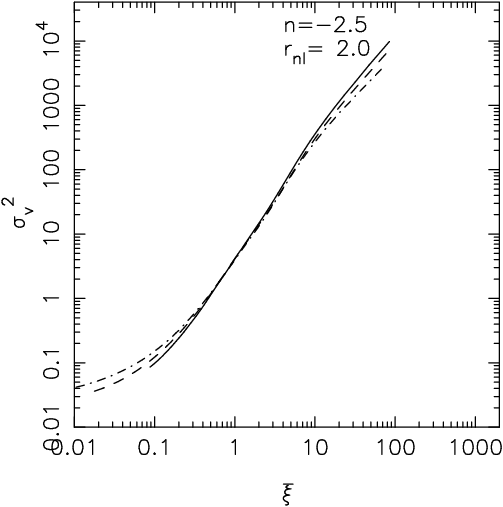} &
\includegraphics[height=2.1truein]{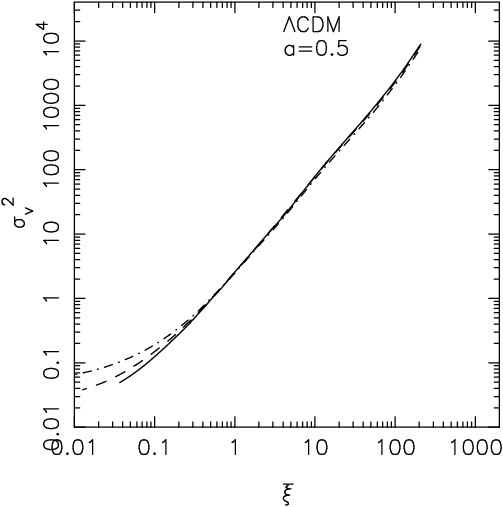} \\
\includegraphics[height=2.1truein]{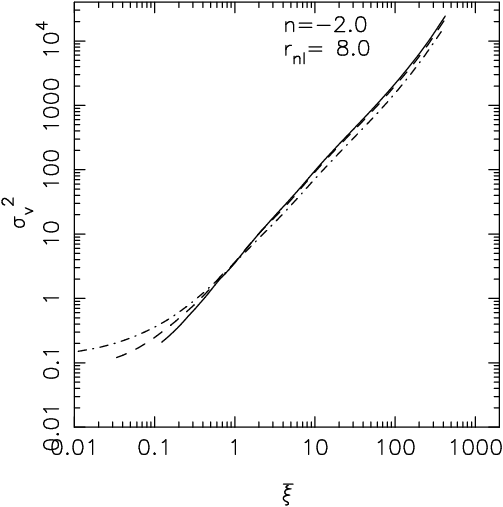} &
\includegraphics[height=2.1truein]{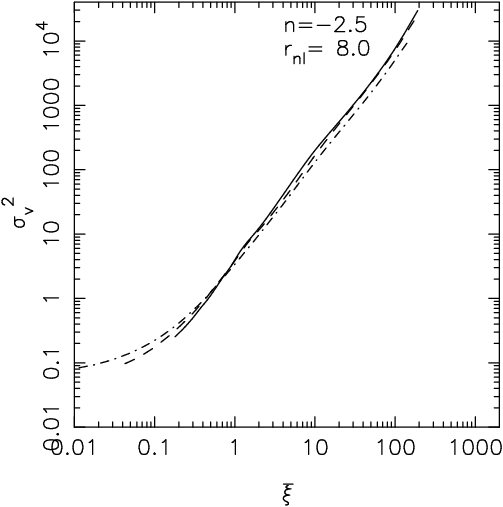} &
\includegraphics[height=2.1truein]{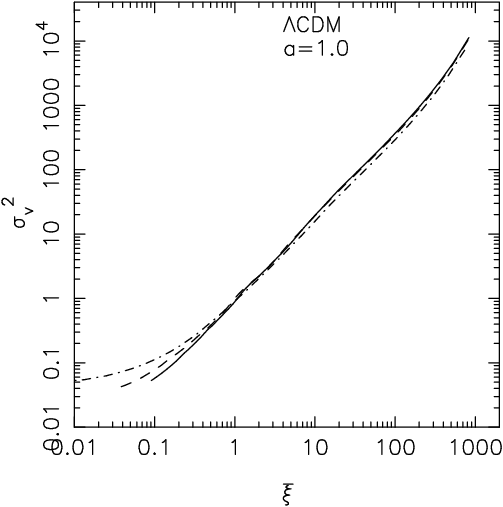} \\
\end{tabular}
\end{center}
\caption{The first two rows show the pair velocity dispersion as a function of
  distance, 
  and the lowest two rows show the pair velocity dispersion as a function of
  average two point correlation function.
The first and third row represent the early epoch and 
second and fourth row represent the later epoch.
In all panels models with $k_c=k_f$ (A1, B1 and C1), $k_c=2k_f$ (A2, B2 and
  C2) and $k_c=4k_f$ (A3, B3 and C3) are represented by the solid, dashed and 
dot-dashed lines respectively.
}
\end{figure*}
\begin{figure*}
\begin{center}
\begin{tabular}{ccc}
\includegraphics[height=2.1truein]{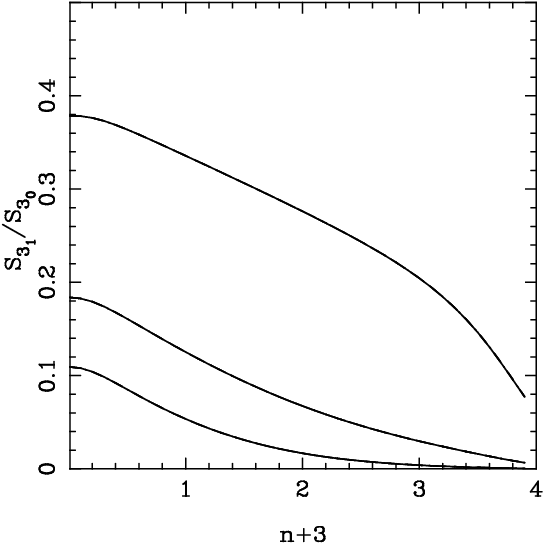} &
\includegraphics[height=2.1truein]{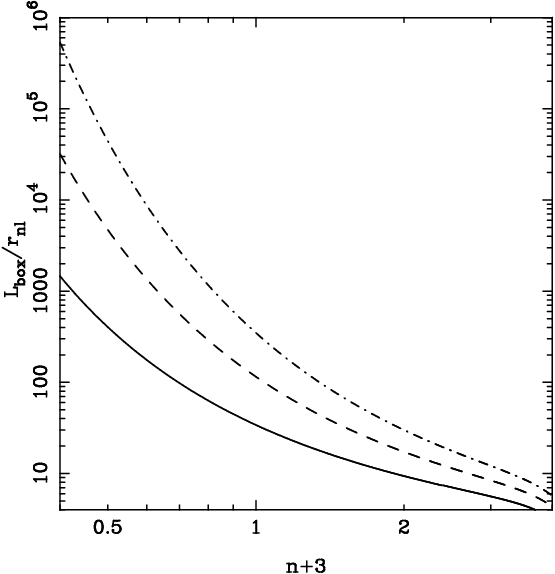} &
\includegraphics[height=2.1truein]{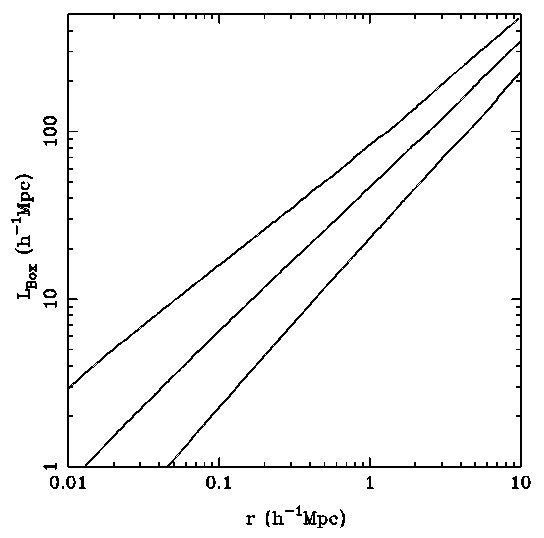} \\
\end{tabular}
\end{center}
\caption{The left panel shows the variation in the fractional correction 
in $S_3$ i.e., $S_{3_1}/S_{3_0}$ (see Eqn.~(4)), with the index of power 
spectrum at the scales $L_{box}/5$ (top curve), $L_{box}/10$ (middle curve)
and $L_{box}/20$ (lowest line). 
For a given tolerance of the error in $S_3$ due to finite box effects, this
gives us the largest scale at which the simulation may be expected to give
reliable results. 
The middle panel shows contours of $C_1$ at the scale of non-linearity
($\sigma_0=1$) for values $C_1=0.01$ (top curve), $0.03$ (middle curve)
and $0.1$ (lower curve).  
The contours are plotted on the $\lbx/r_{nl} ~ - ~ (n+3)$ plane and indicates
the box size required for reliable simulations of a given model.
The right panel shows contours of $S_{3_1}/S_{3_0}$ for the $\Lambda$CDM model
that best fits the WMAP-5 data.  
Contours shown are for $S_{3_1}/S_{3_0} = 0.01$, $0.03$ and $0.1$.  
} 
\end{figure*}

\subsection{Mass Function}

Figure~5 shows the number density of haloes for the three series of
simulations as a function of mass of haloes.  
The haloes have been identified using the Friends of Friends (FOF) method with
a linking length of $0.1$ in units of the grid length.
Plotted in the same panels are the expected values computed using the
Press-Schechter mass function with a correction for the finite box size. 

For each series of simulations, and at each epoch, we fitted the value of
$\delta_c$ to match the simulation with the natural cutoff at the box scale.
The same value of $\delta_c$ is then used for other simulations of the
series. 

We find that the features of the mass function are reproduced correctly by the
analytical approximation, namely: 
\begin{itemize}
\item
The number density of the most massive haloes declines rapidly as the
{\sl effective} box size is reduced.
\item
The number density of low mass haloes increases as the
{\sl effective} box size is reduced.  
This feature is apparent only at the late epoch.
\end{itemize}

\subsection{Velocities}

In our discussion of analytical estimates of the effects of a finite box size
on observable quantities, we have so far omitted any discussion of velocity
statistics. 
The main reason for this is that the power spectrum for velocity is
different as compared to the power spectrum for density and one can get
divergences for quantities analogous to the second order estimators analogous
to $\sigma^2$ for models with $-3 < n \leq -1$.  
This is due to a more significant contribution of long wave modes to the
velocity field than is the case for density.
Relative velocity statistics are more relevant on physical grounds and we use
these for an empirical study of the effects of a finite box size on
velocities. 
It is also important to check whether considerations related to velocity
statistics put a stronger constraint on the box size required for simulations
of a given model.

We measure the radial pair velocity and also the pair velocity dispersion in
the simulations used in this work. 
These quantities are defined as follows:
\begin{equation}
h(r) = -\frac{\langle \left(\mathbf{v}_j -
    \mathbf{v}_i\right).\left(\mathbf{r}_j - \mathbf{r}_i\right)
  \rangle}{aHr_{ij}^2}  
\end{equation}
where the averaging is done over all pairs of particles with separation
$r_{ij}=\left|\left(\mathbf{r}_j - \mathbf{r}_i\right)\right| = r$. 
In practice this is done in a narrow bin in $r$.  
Here $a$ is the scale factor, $H$ is the Hubble parameter and
$\mathbf{v}_i$ is the velocity of the $i$th particle. 
Similarly, the relative pair velocity dispersion is defined as: 
\begin{equation}
\sigma_v^2(r)  = \frac{\langle\left|\mathbf{v}_{ij}\right|^2\rangle }{a^2 H^2
  r_{ij}^2} 
\end{equation}
where $\mathbf{v}_{ij}$ is the relative velocity for a pair of particles, and
averaging is done over pairs with separation $r$. 
Dividing by $a^2 H^2 r_{ij}^2$ gives us a dimensionless quantity and the
usefulness of this is apparent from the following discussion. 

We have plotted the radial component of pair velocity as a function of scale
$r$ in the top two rows of Figure~6.  
Panels in these rows show the pair velocity for the different models at early
and late epochs.  
In each panel, we find that the dependence of pair velocity on $r$ is very
sensitive to the small $k$ cutoff used in generating the initial conditions
for the simulation. 
It has been known for some time
\citep{1991ApJ...374L...1H,1994MNRAS.271..976N} that $h$ is an
almost universal function of $\bar\xi$.
This is certainly true in the linear regime where $h = 2 \bar\xi / 3$ for
clustering in an Einstein-de Sitter universe \citep{1980lssu.book.....P}. 
In order to exploit this aspect, and also to check whether the relation
between $h$ and $\bar\xi$ in the weakly non-linear regime is sensitive to the
box size, we plot $h$ as a function of $\bar\xi$ at the same scale in the
last two rows of Figure~6. 
We find that all runs of a series fall along the same line and 
variations induced by the finite box size are small even in the non-linear
regime.  

Figure~7 shows the relative velocity dispersion as a function
of scale (top two rows) and also as a function of $\bar\xi$ (lowest two rows). 
Again, we find that although the relative pair velocity dispersion at a given
scale is sensitive to the size of the simulation box, its dependence on
$\bar\xi$ is not affected by the large scale cutoff.  
Thus we can use estimates of the correction in $\bar\xi$ to get an estimate of
corrections in pair velocity statistics. 

\section{Summary}

The conclusions of this paper may be summarised as follows:
\begin{itemize}
\item
We have extended our formalism for estimating the effects of a finite box size
beyond the second moment of the density field.
We have given explicit expressions for estimating the skewness and kurtosis in
the weakly non-linear regime when a model is simulated in a finite box size.
\item
We have tested the predictions of our formalism by comparing these with the
values of physical quantities in N-Body simulations where the large scale
modes are set to zero without changing the small scale modes.
\item
We find that the formalism makes accurate predictions for the finite box size
effects on the averaged two point correlation function $\bar\xi$ and
skewness. 
\item
We find that the formalism correctly predicts all the features of
the mass function of a model simulated in a finite box size.  
\item
We studied the effects of a finite box size on relative velocities.  
We find that the effects on relative velocities mirror the effects on
$\bar\xi$.
\end{itemize}

It is desirable that in N-Body simulations the intended model is reproduced at
all scales between the resolution of the simulation and a fairly large
fraction of the simulation box.  
The outer scale up to which the model can be reproduced fixes the effective
dynamical range of simulations. 
One would like $S_3$ be within a stated tolerance of the expected value
at this scale.   
We plot $S_{3_1}/S_{3_0}$ for power law models at the scale $\lbx/20$,
 $\lbx/10$ and $\lbx/5$ in the left panel of Figure~8.
These are plotted as a function of $n+3$.
It can be shown that this ratio, as also $\sigma_1/\sigma_0$ are functions of
scale only through the ratio $r/\lbx$. 
We find that $S_{3_1}/S_{3_0}$ is large for large negative $n$ and decreases
monotonically as $n$ increases.
This ratio is smaller than $10\%$ only for $n \geq 0.8$ at $r=\lbx/5$. 
The corresponding number for $r=\lbx/10$ is $n \geq -1.6$, and for $\lbx/20$
is $n \geq -2.8$
Clearly, the effective dynamic range decreases rapidly as $n+3 \longrightarrow
0$.
This highlights the difficulties associated with simulating such models.

Similarly, one would like $\sigma^2$ and $\sigma_0^2$ to be comparable at the
scale of non-linearity. 
From requirements of self similar evolution of power law models in
simulations, we find that at the scale of non-linearity $C_1 \leq 0.03$
is required for the effects of a finite box size to be ignorable.
This gives us a lower bound on $\lbx/r_{nl}$ for any given model. 
The middle panel shows the required $\lbx/r_{nl}$ as a function of $n+3$ for
$C_1=0.01$, $0.03$ and $0.1$.  
Here $C_1$ is the asymptotic value of $\sigma_1^2$ at $r \ll \lbx$ and is a
fairly good approximation at small scales.  
We find that the required $\lbx/r_{nl}$ for $n=-2$ is more than $100$ for
$C_1^2=0.03$ at the scale of non-linearity.  
Thus we need a simulation with $\lbx \geq 10^3$ if we are to probe the
strongly non-linear regime ($\bar\xi \gg 100$) with some degree of
confidence. 
Requirements for models with $n < -2$ are much more stringent, and for models
like $n=-2.5$ even the largest simulations cannot be used to study the
asymptotic regime. 

To put things in context for the favoured cosmological model, the right panel
in Figure~8 shows contours of $S_{3_1}/S_{3_0}$ in the $r-\lbx$ plane for the
$\Lambda$CDM model that best fits the WMAP-5 data \citep{2008arXiv0803.0586D}. 
We find that in order to ensure that the error in skewness is less than $10\%$
at a scale of $10$~h$^{-1}$Mpc, we need a simulation box of more than
$200$~h$^{-1}$Mpc. 
The required box size is much bigger if the tolerance on error in skewness is
smaller. 
This is a very stringent requirement for simulations of the epoch of
reionization where one would like to get the clustering right at the scales of
a few Mpc.

Given that the formalism we have proposed works well when compared with
simulations, and the fact that calculations in this formalism are fairly
straightforward, we would like to urge the cosmological N-Body simulations
community to make use of this formalism.
We would like to request simulators to report the fractional corrections to
the linearly extrapolated amplitude of clustering and the fractional
correction to skewness across the range of scales of interest. 
This will enable users of simulations to assess potential errors arising due
to a finite simulation volume.


\section*{Acknowledgements}

Numerical experiments for this study were carried out at cluster computing
facility in the Harish-Chandra Research Institute
(http://cluster.mri.ernet.in).
This research has made use of NASA's Astrophysics Data System. 
We would like to thank T. Padmanabhan for useful discussions.
We thank the anonymous referee for useful comments and suggestions. 


\label{lastpage}

\end{document}